\documentstyle[twoside,fleqn,espcrc2]{article}

\def\d{\dagger}
\def\be{\begin{equation}}
\def\eq{\end{equation}}
\def\Tr{{\rm \, Tr \!}}

\input boxedeps.tex
\SetRokickiEPSFSpecial  
\HideDisplacementBoxes
\newcommand{\AmS}{{\protect\the\textfont2
  A\kern-.1667em\lower.5ex\hbox{M}\kern-.125emS}}

\title{Transverse Lattice QCD in 2+1 Dimensions}

\author{S. Dalley\address{Department of Applied Mathematics and
        Theoretical Physics, \\ 
        Silver Street, Cambridge CB3 9EW, England}\thanks{Supported
        at Lat96 by The British Royal Society, DAMTP, and Christ's
        College, Cambridge}
        and 
        B. van de Sande\address{Max Planck Institut f\"ur Kernphysik, \\
        Postfach 10.39.80, D-69029 Heidelberg, Germany}}
       
\begin{document}

\begin{abstract}
Following a suggestion due to Bardeen and Pearson, we formulate
an effective light-front Hamiltonian for large-$N$ gauge theory
in $(2+1)$-dimensions. Two space-time dimensions are continuous
and the remaining space dimension is discretised on a lattice.
Eguchi-Kawai reduction to a $(1+1)$-dimensional theory takes place. 
We investigate the string tension and glueball spectrum, comparing
with Euclidean Lattice Monte Carlo data.
\end{abstract}

\maketitle

\section{Transverse Lattices}
A number of years ago Bardeen and Pearson \cite{bard1} 
formulated a light-front Hamiltonian lattice gauge theory,
which  makes use of the fact that two components of the gauge field are 
unphysical. 
In this
approach {\em two} spacetime dimensions are continuous while the remaining
`transverse' spatial dimensions are discretised on a lattice. 
An empirical study \cite{bard2} of
$SU(\infty)$ gauge theory in $(3+1)$-dimensions produced a 
rough glueball spectrum, but results were inconclusive.
In this exploratory study we bring more recent techniques and ideas to
bear on this problem, choosing pure $SU(\infty)$ Yang-Mills theory in 
$(2+1)$-dimensions as a trial. 
We shall unashamedly use as a benchmark the Euclidean Lattice Monte
Carlo (ELMC) glueball results of Teper \cite{teper}. 
We have applied analytic and numerical techniques to measure 
the transverse string tension and glueball spectrum, as well as many
other observables not described here. 

For $2+1$ dimensions one leaves
longitudinal  co-ordinates $x^0$ and $x^2$ and gauge
fields $A_0$ and $A_2$ intact, while making the `transverse' co-ordinate
$x^1$ discrete, introducing a Wilson link variable $U$ on each
transverse link.
Following Ref.\cite{bard1}
we will assume that link variables $U$ below a certain lattice scale have
been
`blocked' to yield an $N \times N$ complex
matrix $M_{x^1}$ on the link between neighboring sites 
$x^1$ and $x^1+a$  of a sub-lattice. This is reminiscent of a
Dielectric Lattice Gauge Theory \cite{mack}, 
where the magnitude of $M$ plays the
role of dielectric constant. In the confining phase, $M=0$ should be the
prefered vacuum solution. The effective potential obtained through blocking
is difficult to derive analytically, so we will model it
in this work. We choose a Lagrangian density up to 4th order in link
fields
\begin{eqnarray}
{\cal L} & =& 
+ {1 \over 2 a^2 g^2} \Tr\left\{ D_{\alpha} M_{x^1}D^{\alpha} 
M_{x^1}^{\d}\right\} \nonumber \\
&&- {1 \over 4g^2} \Tr\left\{F_{\alpha\beta}F^{\alpha\beta}\right\}  
- V_{x^1}(M)
\label{lag}
\end{eqnarray}
where
\begin{eqnarray}
  V_{x^1}(M) & = & \mu^2  \Tr \left\{M_{x^1}M_{x^1}^{\d}\right\}
\nonumber \\
&& + {\lambda_1 \over a N}  \Tr \left\{M_{x^1}M_{x^1}^{\d} 
   M_{x^1}M_{x^1}^{\d} \right\}  \\
&&+ {\lambda_2 \over a N}  \Tr \left\{M_{x^1}M_{x^1}^{\d} 
   M_{x^1-a}^{\d}M_{x^1-a}\right\} \nonumber
\label{pot}
\end{eqnarray}
and
\begin{eqnarray}
D_{\alpha} M_{x^1}&  = & \left[\partial_{\alpha} +i A_{\alpha} (x^1)\right]
        M_{x^1} \nonumber \\
&&  -  iM_{x^1} A_{\alpha}(x^1+a) 
\label{covdiv}
\end{eqnarray}
$\alpha, \beta \in \{x^0,x^2\}$. 
${\cal L}$ reduces to the usual $2+1$ Yang Mills 
density if $V$ is tuned so that $M \to U/\sqrt{2ag^2}$ as $a \to
0$. This would seem to require (eventually) $\mu^2 < 0$. Since our
quantisation is restricted to $\mu^2 >0$, we should access continuum
physics by improving the action/operators. In particular, we
implicitly assume (for now) that higher order terms in (\ref{lag}) get
progressively smaller (e.g. on dimensional grounds).

In light-front coordinates $x^\pm = (x^0 \pm x^2)/\sqrt{2}$, we take
$x^+$ as canonical time and choose the light-front gauge
$A_- =0$. The theory has a conserved current
\be
  J^{+}_{x^1}
  =  i [M_{x^1} \stackrel{\leftrightarrow}{\partial_{-}} 
M_{x^1}^{\d}  + M_{x^1-a}^{\d} 
\stackrel{\leftrightarrow}{\partial_{-}} M_{x^1-a} ] 
\eq
at each transverse lattice site $x^1$.  
The field $A_+$ is constrained, obeying
$\partial_{-}^{2} A_+ = g^2 J^+ /a$ at each site.  
Solving this constraint leaves only physical fields, in terms of which
the light-front momentum and energy are
\begin{eqnarray}
  P^+ \! \! & \! = \! & \! \! 
2 \! \int dx^- \sum_{x^1} \Tr  \left\{ \partial_- M_{x^1} 
\partial_- M_{x^1}^{\d} \right\} \nonumber \\
 P^- \! & \! = \! & \!
  \int dx^- \sum_{x^1} V_{x^1}(M) - {g^2 \over 2a} \Tr \left\{ 
      J^{+}_{x^1} \frac{1}{\partial_{-}^{2}} J^{+}_{x^1}
           \right\}\nonumber
\end{eqnarray}
There remains residual gauge symmetry under $x^-$-independent
transformations at each site $x^1$.
The zero mode of the $A_+$ constraint equation forces the
corresponding
charge to zero, $\int dx^- J^{+}_{x^1} = 0$. 
This gives a Hilbert space at fixed $x^+$ formed from 
all possible closed Wilson loops of link
matrices $M$ on the transverse lattice
(the $x^-$ co-ordinate of each link field remains arbitrary). Note that
at $N=\infty$  we do not distinguish
$U(N)$ from $SU(N)$, and the effective gauge coupling is $g^2 N$. 
Also, since the loop-loop coupling constant is $1/N$, 
we need not include any more than single loops in the Hilbert space.
In fact when $P^1 = 0$ 
we can simply drop the site indices from $M$ and $P^\alpha$ and the 
Eguchi-Kawai large-$N$ 
reduction to a one link transverse lattice becomes apparent~\cite{dalley1}.
That is, in this frame the theory is isomorphic to one defined on
a one-link transverse lattice 
with periodic boundary conditions, where $P^\alpha$ acts on
a basis of zero winding number loops.

It is convenient to work in longitudinal momentum space at $x^+ =0$
\begin{eqnarray}
 M_{ij}(x^-) & = & 
\frac{1}{\sqrt{4 \pi }} \int_{0}^{\infty} {dk \over {\sqrt k}}
    \  \{a_{-1,ij}(k) e^{ -i k x^-} \nonumber \\ && \ \ \ \ \ \ \ + \
    (a_{+1,ji}(k))^{\d} e^{ i k x^-} \}
\end{eqnarray}
where the modes satisfy equal-$x^+$ commutators
\begin{eqnarray}
\left[a_{\lambda,ij}(k), (a_{\rho,kl}(\tilde{k}))^{\d}\right] 
& = & \delta_{ik} \delta_{jl} \delta_{\lambda \rho}
\delta(k-\tilde{k}) \\
\left[a_{\l,ij}(k),a_{\rho,kl}(\tilde{k})\right] & = & 0
\end{eqnarray}
In the last two expressions $i,j \in \{1,\ldots,N\}$, $\lambda, 
\rho \in
\{+1,-1\}$, and
$\left(a_{\l,ij}(k)\right)^{\d} = (a^{\d}_{\l}(k))_{ji}$.
$a_{\pm 1}^{\d}$ creates a
link mode with orientation $\pm$ on the lattice. 
The  Eguchi-Kawai reduced states corresponding
to $P^1 =0$ and fixed $P^+$ can be written as linear
combinations of singlet Fock basis states  
(summation on repeated indices implied)
\begin{eqnarray}
 & &  \sum_{\begin{array}{@{}c@{}}
    \scriptstyle p=|n|,|n|+2,\ldots \\
    \scriptstyle p>0\end{array}} 
\! \! \int_{0}^{P^+} {dk_1 \cdots dk_p\over N^{p/2}} 
  \nonumber \\
&&\cdot  \delta \! \left(P^+ - \sum_{m=1}^p k_m\right) 
f^{\lambda \rho \ldots \sigma}(k_1,\ldots,k_p) 
 \nonumber \\
&& \cdot \Tr \left\{ a_{\lambda}^{\d}(k_1) a_{\rho}^{\d}(k_2)\cdots
a_{\sigma}^{\d}(k_p)\right\} \,  |0\rangle 
\label{wf}
\end{eqnarray}
where we set the winding number $n=  \lambda + \rho + \cdots + \sigma$ 
equal to zero. 
It remains to
find the coefficient functions $f$, cyclically symmetric in their
arguments,
 which diagonalise  the Hamiltonian $P^-$ and hence
the $(\mbox{mass})^2$ operator $2P^+ P^-$ with eigenvalue $M^2$.

\section{The Boundstate Problem.}

The renormalisation of the quantum theory follows that of a 2D gauge
theory with adjoint matter, involving only  self-energy
correction to the propagator through normal ordering of
interactions~\cite{AD}. 
To diagonalise $P^-$ we  employed both
the analytic method of using an ansatz for the $f$'s and the
numerical one of discretising the momenta $k$ (DLCQ~\cite{brod}).

The theory possesses several discrete symmetries. Charge conjugation
induces the symmetry ${\cal C}: \, a_{+1,ij}^{\d} \leftrightarrow 
a_{-1,ji}^{\d}$. There are two orthogonal reflection symmetries ${\cal
P}_1$
and ${\cal P}_2$ either of which may be used as `parity.'\@ 
If ${\cal P}_1:  x^1 \to -x^1$,  we have 
${\cal P}_1:  \, a_{+1,ij}^{\d} \leftrightarrow a_{-1,ij}^{\d}$. 
${\cal P}_2$: $x^2 \to -x^2$, is complicated in light-front
formalism. Its explicit operation is known only for free
particles, which we call ``Hornbostel parity.''\@  The latter is
nevertheless useful since it is often an approximate quantum number
and its expectation value can be used to estimate ${\cal P}_2$ 
\cite{AD,bvds}.
Given ${\cal P}_2$ and ${\cal P}_1$ we can determine whether spin
${\cal J}$ is even or odd using the relation 
$(-1)^{{\cal J}} = {\cal P}_1 {\cal P}_2$.
If rotational symmetry has been restored in the theory, 
states of spin ${\cal J} \neq 0$ should 
form degenerate ${\cal P}_1$ doublets 
$|+{\cal J}\rangle \pm |-{\cal J}\rangle$~\cite{teper}.
We use ``spectroscopic notation'' 
$|{\cal J}|^{{\cal P}_1 {\cal C}}$ to classify states.

The parameters $\mu^2$, $\lambda_1$, and $\lambda_2$ 
of the effective potential $V(M)$ are unknown functions
of the dimensionless parameter $ag^2N = 1/\beta$, while $g^2N$ should set the
overall
mass scale. Empirically they can be fit to a spectrum, then some 
other quantity of interest predicted (e.g. structure functions), 
or be fixed by examining Lorentz invariance.
To measure the 
string tension in the $x^1$ direction, we
consider a lattice with $n$ transverse links and periodic
boundary conditions.  We construct a basis of Polyakov loops
or ``winding modes'' that wind once around this lattice
and calculate the lowest eigenvalue $M^2$. 
Because of Eguchi-Kawai reduction, this is equivalent to using
Polyakov
loops of winding number $n$ on the single-link periodic lattice
(Eq.\ (\ref{wf}) with $n \neq 0$).
$\Delta (M)/\Delta(n)$ measures the bare string tension.
While this vanishing would signal restoration of translation
invariance, it by no means 
ensures rotational invariance since we have treated the
action anisotropically. We can attempt to ensure rotational invariance
by forming parity doublets in our glueball spectrum for example.

The behavior of the coefficient functions $f$ in Eq.\ (\ref{wf})
when any one of the arguments vanishes is 
\begin{eqnarray}
& \lim_{k_1 \to 0}
 f_{\lambda, \rho, \ldots ,\sigma}(k_1, k_2, \ldots , k_p) \propto k_{1}^{s}&
\nonumber \\  & {2g^2 N \over \pi a} \pi s\tan{(\pi s)} = \mu^2 
&  \label{vanish}
\end{eqnarray}
Specifying also the number of nodes of $f$ as
a function of momenta,
one can make a sensible ansatz. To a first approximation (generic $V(M)$) an 
eigenstate (\ref{wf}) has predominantly a fixed number of link fields
$p$, the mass increasing with $p$ due to the mass term $\mu^2$ in $P^{-}$.
For a given $p$, the energy also tends to increase with the number of
nodes in the wavefunction $f$ due to the
$\tilde{J}(k)\tilde{J}(-k)/k^2$ term.
Thus one expects the lowest
two glueball eigenstates to be approximately
\begin{eqnarray}
&& \int_{0}^{P^+} \! dk  \, f_{+1,-1}(k,P^{+} - k) \nonumber \\
&& \ \ \ \ \ \ \
\cdot  \Tr \left\{ a_{+1}^{\d}(k) a_{-1}^{\d}(P^+-k) \right\} |0\rangle
\end{eqnarray}
with the lowest state having positive symmetric $f_{+1,-1}(k,P^+-k)$,
corresponding to $0^{++}$, 
and first excited state having $f_{+1,-1}$ antisymmetric with one zero, 
corresponding to $0^{--}$. 
The next highest states are either a
4-link state with positive symmetric wavefunctions $f_{+1,+1,-1,-1}$
and  $f_{+1,-1,+1,-1}$ or a symmetric
2-link state with $f_{+1,-1}$ having two zeros. In the glueball spectrum
we  identify the latter states as $0^{++}_{*}$ and $2^{++}$,
although actual eigenstates are a mixture of these.


In our numerical solutions we restrict the number of link fields in our basis
(\ref{wf}) to be $p \leq p_{\rm max}$ and discretise momenta
by demanding antiperiodicity of the fields in $x^- \to x^-+L$.
For fixed integer valued cut-off $K= L P^+/(2\pi)$ momenta
are labeled by odd half integers $\kappa_m = K k_m/P^+$, 
$\sum_m \kappa_m = K$. 
We diagonalise $P^-$
on a computer and 
study the system as $p_{\rm max} \to \infty$ and $K \to \infty$.
At fixed $(p_{\rm max},K)$ we swept the coupling constant space of $V(M)$
and show here some results for the  string tension and glueball spectrum.


\begin{figure}
\BoxedEPSF{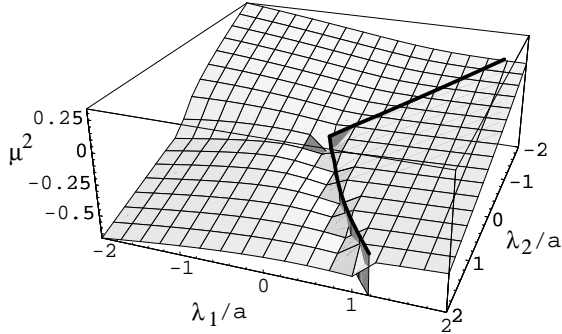 scaled 900}
\caption{ 
Parameters such that the lowest $M^2$
eigenvalues are equal for $n=4$ and $5$ winding modes,
where  $p_{\rm max}=n+4$,
and $K=10.5$ or $11$.  This is an estimate of vanishing bare string
tension.  Also shown is a line such that the $M^2$ eigenvalues are 
approximately degenerate for $n=3$, $4$, and $5$. \label{fig1}}
\end{figure}


\begin{figure}
\BoxedEPSF{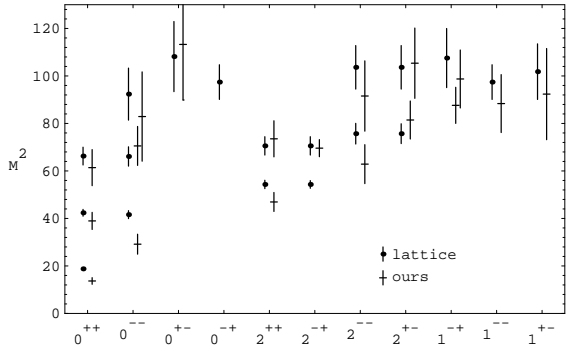 scaled 580}
\caption{
A comparison of our low-lying spectrum with $SU(3)$ ELMC data in units of the 
physical string tension \protect\cite{teper} 
for various $|{\cal J}|^{{\cal P}_1 {\cal C}}$.  
The parameters $g^2 N/a=3.44$, $a\mu^2=0.2 g^2 N$,
$\lambda_1 = 0.34 g^2 N$, and $\lambda_2 = 1.27 g^2 N$ were chosen by
a best fit to the lattice data, $\chi^2 = 45$, where $p_{\rm max}=6$ 
and $K=14$.
Our error estimates are solely for the 
purpose of performing the $\chi^2$ fit.  
\label{fig2}}
\end{figure}

In general  $M^2$ vs.\ $n$ plots for winding modes show 
a good fit to the form $M^2 = An^2 -B$, in
agreement with the expectations of string theory. 
Taking into account all information we have gathered, an acceptable
theory occurs only in the ``wedge shaped region'' 
$-\lambda_1 \leq \lambda_2 \leq \lambda_1/2$.
The renormalised trajectory is most likely to pass through decreasing
$\mu^2$
at $\lambda_2 >0$.

In the glueball spectrum Fig.~\ref{fig2} we 
label the lowest $2^{--}$ and second $0^{--}$ states based on 
$<p>$.  We determine  
$(-1)^{\cal J}$ based on Hornbostel parity;
the exception is the $\left|{\cal J}\right|^{+-}$ sector where
Hornbostel parity gave exactly the opposite of the desired (i.e.
Teper's) results. 
Although Fig.~\ref{fig2} 
indicates that
qualitative agreement can be obtained with the 
ELMC data, there is alarming discrepancy from the expected
degenerate parity doublet $2^{\pm +}$. 
This discrepancy is responsible for almost all of the $\chi^2$
error in our fitting procedure.
There are undoubtedly errors associated with $K$ and $p_{\rm max}$,
but we do not feel that they are  sufficient to account for the 
differences.  In fact, we have examined spectra for $K=10$ and 
$p_{\rm max}=4,6,8$, extrapolating to large $p_{\rm max}$,
along with $p_{\rm max}=6$ and  $K=10,11,12,13,14$, extrapolating
to large $K$. Comparing  with large $N$ extrapolated ELMC
spectra, in either case we saw no great improvement in our results.
There is another quartic term we could have added,
${1\over N^2} (\Tr \ M^{\dagger} M)^2$, which gives non-zero
contribution only on the link---anti-link Fock state. It improves the parity
degeneracies at the expense of a less good `radial' excitation spectrum.

Clearly it is necessary to check the effect of higher order terms in
the effective potential $V(M)$ to see if they are small and capable of
accounting for the discrepancies between our results and Teper's.
If agreement can be obtained in the clean environment of pure glue in
$2+1$ dimensions, we see no reason why the same methods cannot be
applied in practice to spectra, form factors, and structure functions in $3+1$
dimensions.

\vspace{5mm}
\noindent Acknowledgements: We thank M.Burkardt,  H-J.Pirner, and
M.Teper for discussions, and SD thanks H-C.Pauli for hospitality at
the MPI Heidelberg.

\end{document}